# Comparative study of a [WC 6] nucleus with other emission-lines nuclei of planetary nebulae ⋆

A. Acker[1], S. K. Górny[1,2], and F. Cuisinier[1,3]

[1] URA 1280, Equipe 'évolution galactique', Observatoire de Strasbourg, 11 rue de l'Université, F–67000 Strasbourg, France

[2] Laboratory for Astrophysics, Copernicus Astronomical Center, Rabianska 8, PL-87100 Torun, Poland

[3] Astronomisches Institut der Universitat Basel, Venusstrasse 7, CH-4102 Binningen, Switzerland

**Abstract.** The central star (CSPN) of the planetary nebula M 1-25 (PN G 4.9+4.9) is classified as a [WC 6] star, the only CSPN of this subclass known at this time. A comparison with the other emission-lines CSPN (the [WC]-class and the 'weak emission-lines stars' or *wels*-class) shows that (1) the characteristics of this [WC 6] star fit well inside the main properties of the other [WC] CSPN ; (2) the [WC] CSPN seem to evolve from the [WC 8-11] (latter) to [WC 2-4] (earlier) subclasses, say from dense nebulae with cool stars to more extended nebulae with hot nuclei, as for other CSPN ; (3) on the two-colour IRAS diagram, the [WC] and the *wels* CSPN form two different groups : the progenitors of the [WC]-type CSPN seem to be Carbon stars evolving along post-AGB tracks, whereas the *wels* CSPN seem to be related to OH/IR stars, some of them having possibly experienced a late helium-flash.

# 1. Introduction

Planetary nebulae (PN) represent a short transition phase between the asymptotic giant branch (AGB) and the white dwarf domain. About 1150 PN are known in our Galaxy (from the Strasbourg-ESO catalogue, Acker et al., 1992), about 350 of which show a visible central star (CSPN) spectrum. Stellar emission-lines have been detected for about 30% of all known CSPN, among them 47 are classified as [WC]-type CSPN, 38 are assigned to a 'weak emission-lines star' CSPN (*wels*) class, 2 are WN8 stars belonging most probably to pop. I WR stars (Tylenda et al., 1993), and about 15 are of uncertain class. The [WC] nuclei mostly populate subclasses [WC 3-4] and [WC 9-11] whereas the population I WC stars concentrate within WC 5-9. This means that about 13% of the CSPN pass through the WR phase - 40% of them belonging to [WC 3-4] types - and 12% pass through the *wels* phase.

The [WC] stars, considered as the low-mass counterparts of the WR stars, are very likely helium burning, which evolve either along post-AGB helium-burning tracks, or after a late thermal pulse which had occured during the post-PN cooling phase and pushed the star back to the PN regime. At the present time, it is not possible to decide, for a given PN, between the two scenarii, or whether the [WC]-type CSPN represent a phase of binary systems ?

Another question: is the apparent lack of [WC 5-7] subclasses related to the time-scale of one of these scenarii ?

In this paper, we confirm the [WC 6] classification of the central star of M 1-25. We compare the properties of this peculiar (?) nucleus to the properties of the other [WC] and *wels* nuclei. This study allows us to propose that the classification into [WC] subclasses, or into *wels* type, is related to different evolutionary sequences.

# 2. Observations

In the framework of the spectrophotometric survey of planetary nebulae, a spectrum of M 1-25 was taken by Stenholm in July 1986 with the 1.5 m telescope of the ESO at La Silla, using the B&C spectrograph equipped with the IDS device, at low resolution and with an exposure time of 10 min (see figure 1 in Tylenda et al., 1993). Despite the limited quality of the spectrum, we could classify the CSPN as being of "WC 6 ?" type.

New spectra of M 1-25 and some other WC-type CSPN (see Table 1) were taken -

completed with spectra of spectrophotometric standard stars, as usually for calibration purposes - by F. Cuisinier in June 1993 with the same instrument, but equipped with a 2048 x 2048 pixel CCD (360 nm to 740 nm, spectral resolution of $\sim$ 0.4 nm, exposure times of 30 and 60 min, S/N of $\sim$ 10). The reduction and measurements were done with MIDAS facilities. Fig. 1 shows, as an example, the central part of the spectrum of M 1-25.

Also, we obtained observations of northern CSPN in December 1993 and January 1994

Table 1. New [WC] classification for CSPN

| PNG number | Usual name | [WC] class | (our previous class) |
|---|---|---|---|
| 004.9+04.9 | M 1−25 | WC 6 | (WC 6 ?) |
| 019.7−04.5 | M 1−60 | WC 4 | (WC 4−6) |
| 089.0+00.3 | NGC 7026 | WC 3/4 | (WC) |
| 130.2+01.3 | IC 1747 | WC 4 | (WC) |
| 278.8+04.9 | PB 6 | WC 3/OVI | (WC 3 ?) |
| 336.2−06.9 | PC 14 | WC 4 | (WC 4−6) |

at the 1.52 m telescope at the Haute-Provence Observatory equipped with the Aurélie spectrograph (510 nm to 600 nm, spectral resolution of $\sim$ 0.1 nm, exposure times of 1 to 4 hours, S/N of 10 to 50, see description in Gillet et al., 1994).

3. Comparative analysis

*3.1. Classification*

The quality of the new spectra allows us to refine the classification of some [WC] CSPN, as shown on Table 1, by using the criteria defined by Mendez & Niemela (1982). In particular, we confirm that the nucleus of M 1-25 is of [WC 6]-subclass:

– the ratio CIV (580nm) / CIII (569.5) = 16.0 ;

– CIII (569.5) / OV (559.5) = 1.1 ;

– OVI (529) is practically invisible (see Fig. 1).

No other object was found to lie in the [WC 5-7] gap.

Fig. 2 shows the distribution of CSPN along [WC] subclasses. CSPN of uncertain status (subclass [WC 4-6], or [WC] in Tylenda et al., 1993) are not shown. The subclass [WC 10] is apparently absent, probably due to a definition of [WC 10-12] subclasses (Hu & Bibo, 1990) without link to the earlier subclasses. Objects classified as [WC 11] by Hu & Bibo were classified as [WC 10] by Mendez & Niemela (1982). Therefore we propose to change the [WC 11] subclass into [WC 10-11] subclass.

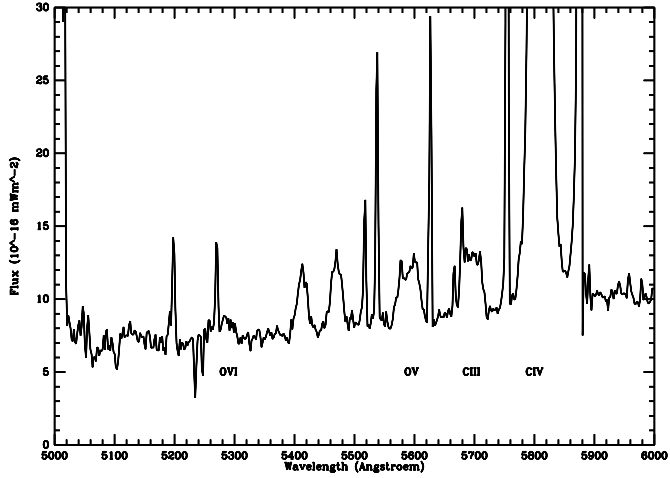

**Fig. 1.** Spectrum of M 1-25 (central part), taken with the 1.5 m telescope of the ESO at La Silla, using the B&C spectrograph equipped with a 2048 pixels CCD (exposure time : 60 min)

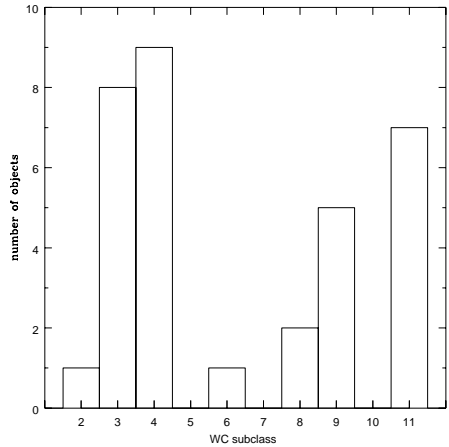

**Fig. 2.** Histogram of objects in a given [WC] subclass. Uncertain class are not shown (two WC 3-4, three WC 4-5 and six WC 4-6)

## 4. Distribution

If the [WC] CSPN come from stars originally more massive, they should be confined closer to the galactic disk when compared to other PN. A Kolmogorov-Smirnov test shows no significative difference of the distributions in galactic coordinates of [WC] CSPN and other PN. In addition, within $|b| < 7^o$, one finds 77 % of the [WC] CSPN, and 73 % of all PN. For all objects belonging probably to the galactic bulge (|l| and |b| smaller than 7, diameter < 10 "), we find 14 % of the [WC] CSPN, and 18 % of all PN. All these observed similarities in the spatial distribution indicate that the distribution of the progenitors masses is also similar, if we assume a unique Mi-Mf relation.

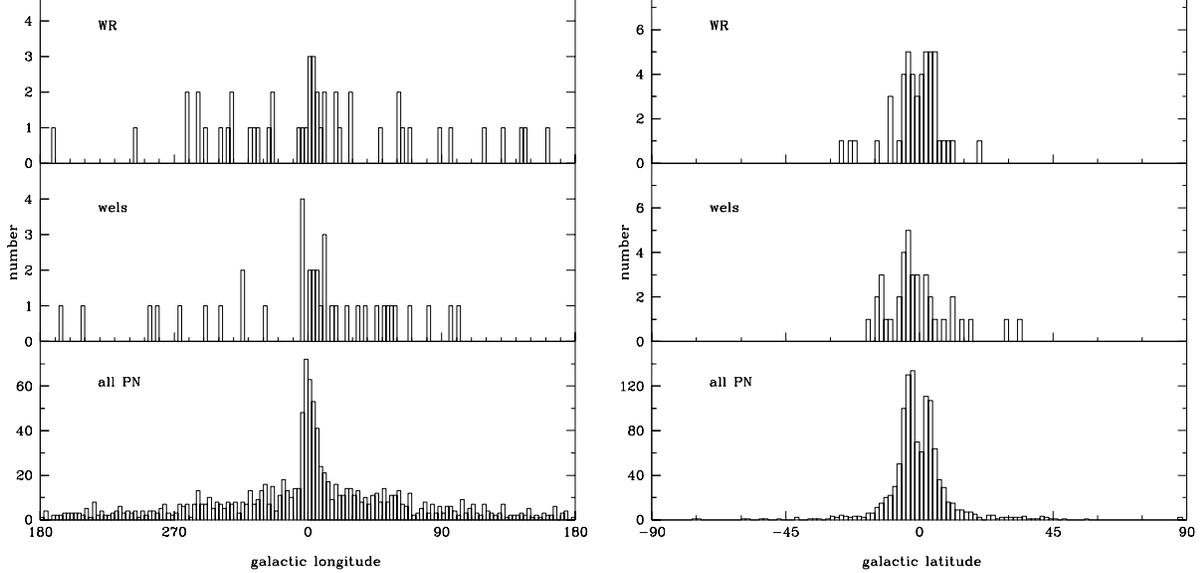

**Fig. 3.** Distribution of [WC] and *wels* PN, in galactic coordinates

Fig. 3 shows that the same remark should be applied to the *wels* CSPN, too. The same conclusion could be done from the distribution of the radial velocity along the galactic longitude (Fig. 4): kinematical behaviour of all three samples is very similar.

*4.1. Physical parameters*

**M 1-25.** From our 1993 spectra, and from high resolution spectra taken with the 1.4 m telescope at ESO in the framework of an other program devoted to expansion velocities of the galactic bulge PN, see Acker, 1993), we have calculated the following parameters for M 1-25 :

– T of the central star (Zanstra method) : $\log T_z(He) = 4.78$

– Expansion velocity : $V_{H_\alpha} = 35$ km/s, V[NII] = 39 km/s, V[OIII] = 35 km/s

– Electron temperature : 7800 K (using the [OIII] line ratio)

– Electron density : $6800/cm^3$ (using the [SII] doublet ratio)

**Other [WC] CSPN.** The **stellar temperature**, calculated for most CSPN through the Zanstra method using the data listed in Tylenda, Stasińska, et al., 1994, and using in other cases the Energy-Balance method (Preite-Martinez et al., 1991), rises continuously, as expected, from latter to earlier subclasses (Fig. 5). This graph does not show any significative deficiency of CSPN having middle $T_z$ as would have been expected from the observed lack of [WC 5-7] subclasses.

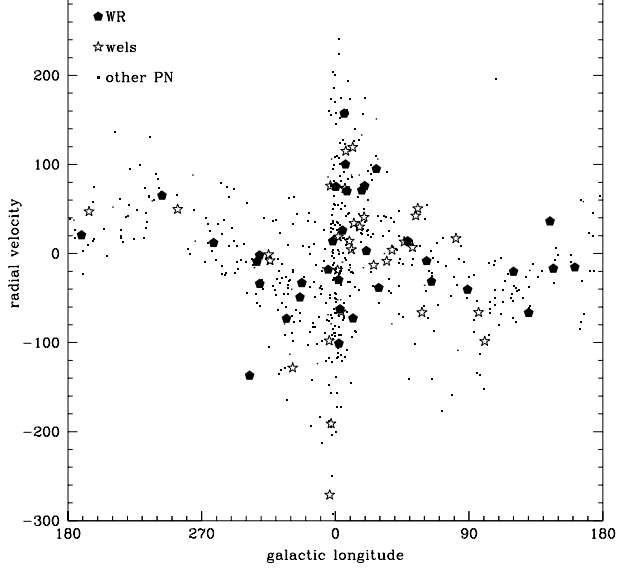

**Fig. 4.** Radial velocity versus galactic longitudes of [WC] and *wels*

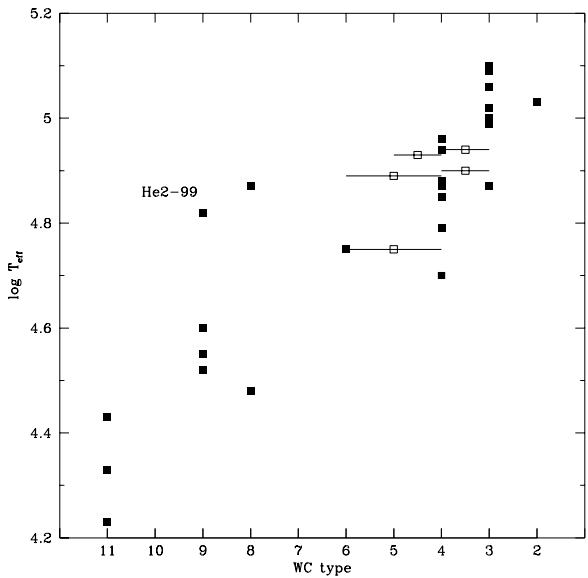

**Fig. 5.** Temperature of the star versus the [WC] subclass. Uncertain classes are shown by empty squares.

The **expansion velocity** (taken from the catalogue by Acker et al., 1992) increases from cool [WC] CSPN to the hottest one, with a wide range of values for the late types (as already shown in Górny & Tylenda, 1993). An opposite trend is shown for the **nebular** low density show higher expansion velocities than dense, compact PN. As the nebular density decreases in the course of time (see Schmidt-Voigt & Köppen, 1987), we should conclude that the [WC] CSPN evolve from latter to earlier subclasses, or from dense nebulae with cool stars to more extended nebulae with hot nuclei, as for other CSPN.

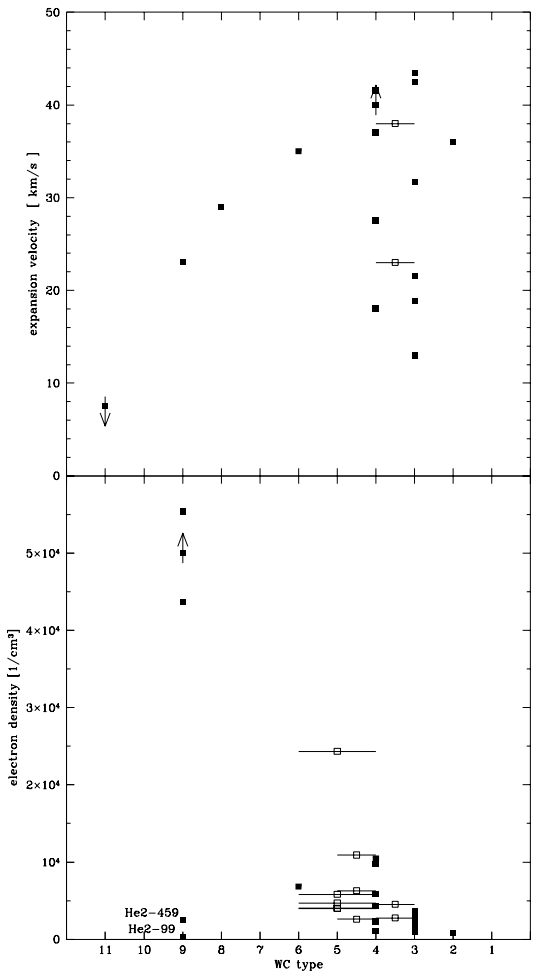

**Fig. 6.** Expansion velocity of the nebulae and electron density of the nebulae versus the [WC] subclass

Table 2 presents the mean values of some parameters : the stellar temperature (col. 2), the expansion velocity (col. 3), the electron density (col. 4), the total infrared luminosity FIR (col. 5) and the infrared excess IRE (col. 6) (values taken from Zhang & Kwok, 1993), for different samples : [WC] CSPN, *wels* CSPN, and H-rich CSPN (as defined by Mendez, 1991). The mean value and the standard deviation are given, the number of objects used to calculate the mean value being indicated in parentheses.

Comparison of the [WC] and *wels* samples (Fig. 7 and Table 2) leads to some remarks:

**Table 2.**

**Mean values of physical parameters for CSPN**

The mean FIR values are calculated without :

(a) NGC 5315 (FIR = 1163, IRE = 2.8) ; NGC 6369 (FIR = 1198, IRE = 1.0). If these objects are included, <FIR> = 281±336 (19)

(b) BD+30°3639 (FIR = 4396, IRE = 7.8). If this object is included, <FIR> = 1144±1480 (7)

(c) The three objects listed in (a) and (b). If these objects are included, <FIR> =513±871 (27)

(d) NGC 6543 (FIR = 1879, IRE = 3.5) ; NGC 6572 (FIR = 2412, IRE = 2.8) ; He 2−131 (FIR = 1538, IRE = 5.1).
If these objects are included, <FIR> = 411±591 (28)

| $CSPN$ | $<\log T^*>$ (N) $K$ | <exp. v> (N) $km.s^{-1}$ | <el. dens.> (N) $10^3 cm^{-3}$ | $<FIR>$ (N) $10^{-18} W\, cm^{-2}$ | $<IRE>$ (N) — |
|---|---|---|---|---|---|
| [WC 2-5] | 4.9 ± .1 (21) | 31 ± 10 (14) | 5.1 ± 3.5 (23) | 175 ± 120 (17) (a) | 3.1 ± 2.3 (19) |
| [WC 6] | 4.8 (1) | 35 (1) | 6.8 (1) | 114 (1) | 3.1 (1) |
| [WC 9-11] | 4.5 ± (8) | 26 ± 4 (2) | 25.5± 2.8 (4) | 602 ± 400 (6) (b) | 7.9 ± 2.4 (7) |
| all [WC] | 4.8 ± .2 (30) | 31 ± 10 (17) | 8.0 ± 12.5 (28) | 286 ± 280 (24) (c) | 4.4 ± 3.1 (27) |
| *wels* | 4.7 ± .2 (32) | 16 ± 8 (18) | 6.0 ± 4.5 (24) | 228 ± 220 (25) (d) | 3.8 ± 1.9 (28) |
| H-rich | 4.7 ± .2 (35) | 22 ± 12 (37) | 8.0 ± 12 (17) | 456 ± 702 (27) | 2.4 ± 2.3 (26) |

− For the 2 samples, the stellar temperature shows some correlation with the expansion velocity and with the density

− [WC] CSPN tend to have higher stellar temperature than the *wels* CSPN

− [WC] CSPN tend to situate at higher expansion velocities than the *wels* and the H-rich CSPN.

*4.2. IRAS colour-colour diagram and evolutionary sequences*

The position of the [WC] subclasses into the IRAS colour-colour plane (Fig. 8) shows a trend of the colours becoming redder from [WC 8-11] to [WC 2-4]. This trend should be compared with the post-AGB evolution, calculated in agreement with the Schönberner tracks, for the [WC 9] star BD+30°3639 (Siebenmorgen et al., 1994). This result, extended to other late [WC] CSPN, ties clearly these stars to the post-AGB objects (Zijlstra et al.,

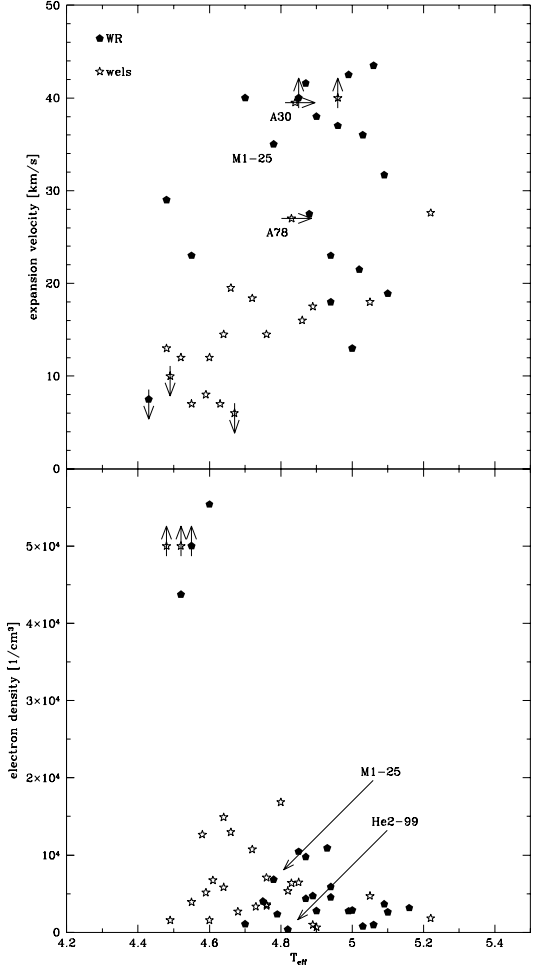

**Fig. 7.** Expansion velocity of the nebulae and electron density of the nebulae versus the temperature of the stars, for [WC] and *wels* PN

1994).

In this scheme, the [WC 8-11] CSPN are young objects, with cool stars, with compact and dense nebulae dominated by dust, being very bright in the IRAS bands and having an IR excess (see values of FIR and IRE in Table 2, taken from Zhang & Kwok, 1993), evolving directly from the AGB. For [WC 3-4] PN, the older nebulae are more extended and cooler, and the ratio $F_{12}/F_{25}$ is lower.

The blue object PC 14 (see Fig. 8) is perhaps more compact than other [WC 4] CSPN, if the expansion is lower in relation with lower metallicity (as for LMC [WC] CSPN studied by Zijlstra et al., 1994) ?

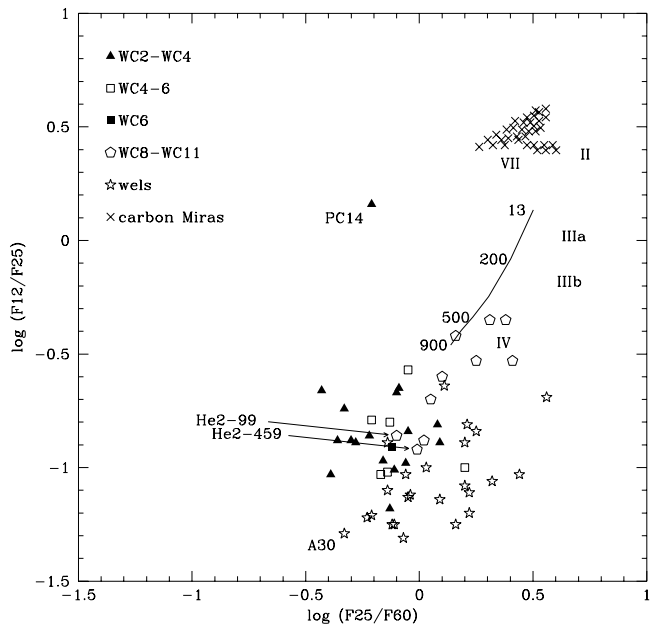

**Fig. 8.** Position of [WC] subclasses and *wels* CSPN in the IRAS two-colour diagram. Position of the [WC 6] CSPN is shown by the black squares.

The curve shows the post-AGB evolution calculated for the progenitor of the [WC 9] star BD+30°3639, where labels designate elapsed time (yrs) after the end of the AGB mass loss (from Siebenmorgen et al., 1994). The crosses indicate the position of the carbon Miras sammples studied by Willems & de Jong (1988), whereas the regions II to VII are reported from van der Veen and Habing (1988) (see the text).

## 5. Conclusion

We can conclude that most of the [WC] CSPN should originate from post-AGB stars, along an evolutionary sequence going from [WC 9-11] (cool stars, dense nebulae) to [WC 2-4] stars (hot stars, extended nebulae), the [WC 5-7] subclass being a short transition phase, with an unique member known at this time, the [WC 6] CSPN of M1-25.

However, some [WC 9-11] CSPN with extended and low density nebulae should more probably be the result of a late helium flash (case of the [WC 9] objects He 2-99 and He 2-459, see Fig. 6 and 8). On the other hand, some old extended PN with very hot nuclei belonging to the *wels* class (A 30, A 58, A 78) should also confirm the second scenario of a late helium flash.

Comparison in the IRAS colour-colour diagram of the [WC] and the *wels* CSPN is shown on Fig. 8. A clear separation appears between [WC] and *wels*, the latter having smaller values by about 0.5 dex of the $\frac{I_{12\mu m}}{I_{25\mu m}}$ ratio. This confirms the analysis in Tylenda et al. (1993), arguing that the [WC] and the *wels* CSPN form probably two distinct groups, the wels having very weaker emission lines, and having also a different chemical composition in the line forming wind : the N/C ratio is on average higher in *wels* than in [WC] CSPN winds. Most of the *wels* stars appear in Mendez (1991) as being O-rich stars. Therefore, the difference of these two groups is certainly related to different chemical abundances. By comparing the position of C- and O-rich stars on the IRAS two-colour diagram (see Habing & Blommaert, 1993 ; van der Veen & Habing, 1988 ; Willems & de Jong, 1988), we may suggest that [WC] CSPN originate from carbon-stars (indicated on Fig. 8 by the region VII defined by van der Veen & Habing, 1988, which contains the carbon Miras samples studied by Willems & de Jong, 1988). The *wels* CSPN are related to variables stars with O-rich circumstellar shells (regions II, III, IV defined by van der Veen & Habing (1988), indicated on Fig. 8). This assumption is compatible with some observed diferences, especially concerning expansion velocities (Fig. 6 and Table 2). As the expansion is a strong function of metallicity, C-rich stars expand faster than O-rich stars, a property observed for LMC OH/IR stars with slower expansion than galactic OH/IR stars, due to the lower metallicity in the LMC and consequently lower dust abundance in the AGB wind (Wood et al., 1992).

**Acknowledgement**

It's a pleasure to thank A. Zijlstra for fruitfull discussions.

S. K. G. acknowledges a support from KBN grant n$^o$. 2-2114-92-03 financed by Polish Commitee for Scientific Research, and from the program " Réseau Formation Recherche" of the French "Ministère de l'Enseignement Supérieur et de la Recherche". He thanks also the Observatoire de Strasbourg for hospitality, and is gratefull to the GdR 968 ("Milieux Circumstellaires").

A. A. acknowledges S. Durand for help in the Latex version.